\documentclass[twocolumn,pre,showpacs]{revtex4}
\usepackage{graphicx,color}
\usepackage{dcolumn}
\usepackage{amsmath,amssymb}

\begin{document}
%
\title{
 Field-induced Berezinskii-Kosterlitz-Thouless transition and
 string-density plateau in the anisotropic triangular antiferromagnetic
 Ising model
}
\author{
 Hiromi Otsuka,$^1$
 Yutaka Okabe,$^1$
 and
 Kouichi Okunishi$^2$
 }
\address{
 $^1$Department of Physics, Tokyo Metropolitan University, Tokyo 192-0397 Japan\\
 $^2$Department of Physics, Niigata University, Niigata 950-2181 Japan
 }
\date{\today}
\begin{abstract}
 The field-induced Berezinskii-Kosterlitz-Thouless (BKT) transition in
 the ground state of the triangular antiferromagnetic Ising model is
 studied by the level-spectroscopy method. 
 We analyze dimensions of operators around the BKT line, and estimate
 the BKT point $H_{\rm c}\simeq0.5229\pm0.001$, which is followed by the
 level-consistency check to demonstrate the accuracy of our estimate.
 Further we investigate the anisotropic case to clarify the stability of
 the field-induced string-density plateau against an incommensurate
 liquid state by the density-matrix renormalization-group method.
\end{abstract}
\maketitle

\begin{figure}[t]
\mbox{\includegraphics[width=2.8in]{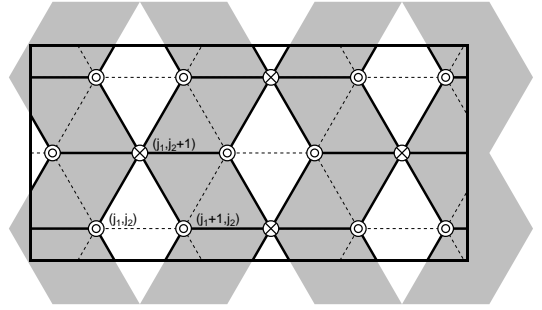}}
 \caption{
 An example spin configuration with the $\sqrt3\times\!\sqrt3$
 structure.
 The $j$th site is specified by two integers $(j_1,j_2)$ as labeled. 
 The long (short) side of the rectangle frame is in the $x_1$ ($x_2$)
 direction. 
 The spins on two (one) of three sublattices $\Lambda_0$ and $\Lambda_1$
 ($\Lambda_{2}$) are parallel $\circledcirc$ (antiparallel $\otimes$) to the
 field direction.
 Four gray stripes in the $x_2$ direction give the string representation
 of the spin configuration.
 }
 \label{FIG1}
\end{figure}

 It was exactly proven that the nearest-neighbor (NN) antiferromagnetic
 Ising model on the triangular lattice shows no phase transition at
 finite temperature, and possesses a ground-state ensemble carrying the
 finite residual entropy per spin, $S\simeq0.3231k_{\rm B}$
 \cite{Wann50}.
 This circumstance stems from the frustration effect that not all spins
 at the corners of each elementary triangle can be energetically
 satisfied, and then brings about the power-law decays of the
 correlation functions of physical quantities in the ground state
 \cite{Step70}. 
 There is a long history of research on various types of
 perturbation effects on this ground-state degeneracy
 \cite{Blot91,Blot93,Kinz81,Qian04a,Qian04,Noh_95},
 where, as we will see in the
 following, the exact mapping to the so-called triangular Ising
 solid-on-solid (TISOS) model
 \cite{Blot82,Nien84}
 followed by a coarse graining provides an
 effective field theory to describe the low-energy physics
 \cite{Jose77}.
 In addition, it should be remarked that the ground-state spin
 configurations can be classified according to the number of strings
 (see below)
 \cite{Blot82},
 which provides an intuitive connection to one-dimensional (1D) quantum
 systems with global U$(1)$ symmetry under the path-integral
 representation.

 In this paper, we treat an anisotropic triangular antiferromagnetic
 Ising model (TAFIM) under a magnetic field; its reduced
 Hamiltonian ${\cal H}=\beta H_{\rm TAFIM}$ is given as
 \begin{equation}
  {\cal H}(K_1,\mu,H)
   = \sum_{\langle j,k\rangle}K_{jk} \delta_{\sigma_j,\sigma_{k}}
   - H \sum_j \delta_{\sigma_j,0}.
   \label{eq_Hamil}
 \end{equation}
 The binary variable $\sigma_j=0,1$ is on the $j$th site of the
 triangular lattice $\Lambda$ which consists of interpenetrating three
 sublattices $\Lambda_l$ $(l=0,1,2)$, and the first (second) sum runs
 over all NN pairs (sites).
 The AF coupling $K_{jk}$ takes two values $K_1+\mu$ or $K_1$ depending
 on whether the bond ${\langle j,k\rangle}$ lies in the $x_1$ direction
 or not (see Fig.\ \ref{FIG1}). 
 Here we define a quantity
 $Q=\sum_{j_1}N_{j_1,j_2}$
 with  
 $N_{j_1,j_2}=1-\delta_{\sigma_{j_1,j_2},\sigma_{j_1+1,j_2}}$ for all
 $j_2$,
 and further restrict ourselves to the zero temperature case
 $K_1\to\infty$. 
 Then $Q$ is independent of $j_2$ and the Boltzmann weight per row is
 given by $e^{\mu Q}$.
 Therefore, the anisotropy parameter $\mu$ plays a role of the
 chemical potential to control the number of strings
 (an example of the string representation is given in Fig.\ \ref{FIG1})
 \cite{Blot82}. 
 Our main goal is to clarify the phase diagram of the model
 (\ref{eq_Hamil}) in its ground state $K_1\to\infty$.
 For this, we shall use the level-spectroscopy (LS) method
 \cite{Nomu95}
 to treat the Berezinskii-Kosterlitz-Thouless (BKT) transition induced
 by $H$ in the isotropic case $\mu=0$.
 On the other hand, for the anisotropic case $\mu\ne0$, the
 Pokrovski-Talapov (PT) transition
 \cite{Pokr79}
 between a commensurate ordered phase and an incommensurate liquid phase
 is expected.
 We directly calculate the $\mu$ dependence of the number of strings by
 the use of the density-matrix renormalization-group (DMRG) method
 \cite{Whit92}.
 Then, we provide a reliable phase diagram.

 As we will see in the following,
 the string degrees of freedom in which the frustration effects are
 encoded play a central role for both understanding the phase diagram
 and relating it to the magnetization plateaus observed in the 1D
 frustrated quantum spin systems 
 \cite{Okun02}.
 This comes from the fact that the quantity $Q$ corresponds to the
 uniform magnetization, and thus $\mu$ can be regarded as the magnetic
 field in the quantum spin systems.
 Further, quite recently, the string representation has been also
 employed in investigations of the 2D frustrated quantum magnets
 \cite{Jian05},
 so the precise analysis on the present fundamental model may also offer
 a hint for a further understanding of quantum systems.

\begin{table}[t]
 \caption{
 Estimates of $H_{\rm c}$. 
 We label the estimate by the spin-wave excitation (the vortex
 excitation) as KT1 (KT2).
 We also give the results by the phenomenological renormalization-group
 (PRG) and the level-spectroscopy (LS) methods.
 } 
 \begin{tabular}{cccrclll}
  \hline\hline
  & Reference     &Method  & Extrapolation    && $H_{\rm c}$& Error& \\ 
  \hline
  & \cite{Blot91} &   KT1  &    A rough estimate && 0.6   & ---       & \\ 
  & \cite{Blot93} &   KT1  &        Iterated fit && 0.532 & $\pm$0.02 & \\ 
  & \cite{Qian04} &   KT2  &        Iterated fit && 0.52  & $\pm$0.04 & \\ 
  & \cite{Quei99} &   PRG  & Finite-size scaling && 0.422 & $\pm$0.014& \\ 
  &  This work    &   LS   &      Polynomial fit && 0.5229& $\pm$0.001& \\ 
  \hline\hline
 \end{tabular}
 \label{TAB_I}
\end{table}

 Now we shall start with the isotropic case $\mu=0$.
 For $H=0$, from the exact asymptotic behavior of the spin-spin
 correlation function the scaling dimension of the staggered
 magnetization ($S$) is given by
 $x_S=\frac14$
 \cite{Step70},
 while the dimension of the uniform magnetization ($s$) is by
 $x_s=\frac94$
 \cite{Nien84}.
 Thus the magnetic field is irrelevant, and the critical region
 continues up to a certain value $H_{\rm c}$.
 For $H>H_{\rm c}$, the criticality of the ensemble disappears and the
 threefold-degenerate ground state with the $\sqrt3\times\!\sqrt3$
 structure of the sublattice is realized, where a majority spin is in
 the field direction  (see Fig.\ \ref{FIG1}).
 The transition at $H_{\rm c}$ is the BKT type, and is described, in the
 scaling limit, by the 2D sine-Gordon Lagrangian density
 \cite{Jose77}
 \begin{equation}
 {\cal L}[\phi]=
  \frac{1}{2\pi K}\left(\nabla \phi\right)^2+
  \frac{y}{2\pi\alpha^2} \cos 3\sqrt2\phi,~~~ K \simeq \frac49, 
  \label{eq_SG}
 \end{equation}
 where $y\propto H$ and the continuous field in the 2D Euclidean space
 $(x_1,x_2)$ proportional to the height variable of the TISOS model
 satisfies $\sqrt2\phi+2\pi=\sqrt2\phi$
 \cite{Comm1}.
 Since the lattice model is not exactly solvable, there have been
 several attempts to numerically estimate $H_{\rm c}$: 
 Bl\"ote and Nightingale investigated this problem in detail by the
 transfer-matrix method
 \cite{Blot91,Blot93}.
 Actually, they evaluated finite-size estimates $H_{\rm c}(L)$ by
 numerically solving the equation for the scaled gap, i.e., the
 so-called KT criterion $x_S(H,L)=\frac29$ (this is referred to as KT1
 in Table\ \ref{TAB_I}).
 Then, in order to accelerate the slow convergence of $H_{\rm c}(L)$,
 the iterated fits with taking account of the logarithmic correction
 were performed (for a more recent estimation, see
 \cite{Qian04}).
 On the other hand, de Queiroz {\it et al.} treated the same model by
 the phenomenological renormalization-group (PRG) method \cite{Quei99};
 they exhibited a much smaller value inconsistent with the previous
 estimations (see Ref.\
 \cite{Qian04}).
 However, it is often pointed out that PRG calculations fail to estimate
 the BKT points
 \cite{Inou99},
 so their result may suffer from an inadequacy of the method.
 Consequently, to accurately determine $H_{\rm c}$, there still remains
 some difficulty.

 In the studies of 1D quantum systems, however, the LS method provides
 an efficient way to treat the BKT transitions
 \cite{Nomu95}.
 This is also true for the investigations of 2D classical spin systems
 \cite{Otsu05}.
 Let us consider the system on $\Lambda$ with
 $M$ ($\to\infty$) rows in the $x_2$ direction of $L$ (a multiple of 3)
 sites in the $x_1$ direction wrapped on the cylinder and define
 the transfer matrix connecting the next-nearest-neighbor rows.
 Since, in our discussion, the number of the strings $Q$ (or its density
 $\rho=Q/L$) is the most important conserved quantity in the transfer,
 we explicitly specify a block of the matrix as ${\bf T}_Q(L)$ and
 denote its eigenvalues as $\lambda_{p,Q}(L)$ or their logarithms as 
 $E_{p,Q}(L)=-\frac12\ln|\lambda_{p,Q}(L)|$ ($p$ specifies a level).  
 In the isotropic case,
 the smallest one corresponding to the ground state is in the
 block $Q_0=2L/3$
 \cite{Blot93,Blot82};
 we shall denote it and the excitation gaps from it as
 $E_{{\rm g},Q_0}(L)$
 and
 $\Delta E_{p,Q}(L)=E_{p,Q}(L)-E_{{\rm g},Q_0}(L)$,
 respectively.
 Then the conformal invariance provides direct expressions for the
 central charge $c$ and a scaling dimension $x_{p,Q}$ in the critical
 system as
 $E_{{\rm g},Q_0}(L)\simeq Lf-\pi c/6L\zeta$
 and
 $\Delta E_{p,Q}(L)\simeq 2\pi x_{p,Q}/L\zeta$.
 Here $\zeta$ $(=2/\sqrt3)$ and $f$ are the geometric factor for
 $\Lambda$ and a free energy per site, respectively
 \cite{Card84,Blot86}.

 Bl\"ote and Nightingale precisely checked various scaling dimensions
 based on the Coulomb-gas scenario
 \cite{Blot93},
 whereas Nomura pointed out the importance of
 logarithmic corrections in the renormalized scaling dimensions
 $x(l)=\Delta E(L)/(2\pi/L\zeta)$ 
 to determine the BKT point $(l=\ln L)$
 \cite{Nomu95}.
 In the present effective theory\ (\ref{eq_SG}), there are two marginal
 operators on the BKT point, i.e., 
 ${\cal M}=(1/K)\left(\nabla \phi\right)^2$
 and 
 $s=\sqrt2\cos3\sqrt2\phi$
 which hybridize along the RG flow and result in two orthogonalized
 ones, i.e.,
 the ``${\mathcal M}$-like''
 and
 the ``cos-like'' operators
 \cite{Nomu95}.
 Writing the former and the latter as
 $O_0\propto\left({\cal M}+s/\sqrt2\right)$ and
 $O_1\propto\left(-{\cal M}/\sqrt2+s\right)$,
 their renormalized scaling dimensions can be calculated near the
 multicritical point; the results up to the first-order perturbations are
 given as follows:
 $x_0(l)\simeq 2-y_0\left(1+\frac43 t\right)$
 and 
 $x_1(l)\simeq 2+y_0\left(2+\frac43 t\right)$,
 where
 $(y_0,y_1)=(9K/2-2,y)$
 and the small deviation from the BKT point
 $t=y_1/y_0-1$
 \cite{Nomu95}.
 Another important operator is a relevant one, i.e.,
 the staggered magnetization $S=\exp(\pm i\sqrt2\phi)$
 whose dimension is expressed as 
 $x_S(l)\simeq \frac29\left(1+\frac12y_0\right)$ 
 in the same region. 
 Consequently, the level-crossing condition,
 \begin{equation}
  x_0(l)=4-9x_S(l),
 \end{equation}
 provides a finite-size estimate of the BKT point. 
 Since these operators are described by $\phi$, we can calculate the
 renormalized scaling dimensions from excitation gaps found in the $Q_0$
 block.
 Further, the symmetry properties such as
 the translation of one lattice spacing (or a cyclic permutation among
 sublattices $\Lambda_i$) ${\cal T}$,
 the space inversion ${\cal P}$,
 and
 the spin reversal ${\cal S}$ at $H=0$
 are also important for the precise specification of relevant excitations
 levels.  
 These symmetry operations can be interpreted in the field language as
 ${\mathcal T}$:\ $\sqrt2\phi\mapsto \sqrt2\phi+2\pi/3$,
 ${\mathcal P}$:\ $\sqrt2\phi\mapsto-\sqrt2\phi       $, and
 ${\mathcal S}$:\ $\sqrt2\phi\mapsto \sqrt2\phi+\pi   $
 \cite{Blot93,Nien84,Land83}.
 For instance, $s$ is invariant for ${\mathcal T}$ and ${\mathcal P}$,
 but is odd for ${\mathcal S}$ as expected, while $S$ is in the
 $|k|=2\pi/3$ block and is odd for ${\mathcal S}$.
 Therefore, the corresponding excitations to $O_{0}$ and $O_{1}$ can be
 found in the subspace of the wavevector $k=0$ and the even parity for
 the space inversion. 
 We calculate the excitation levels by utilizing these symmetry
 operations.

\begin{figure}[t]
\mbox{\includegraphics[width=2.8in]{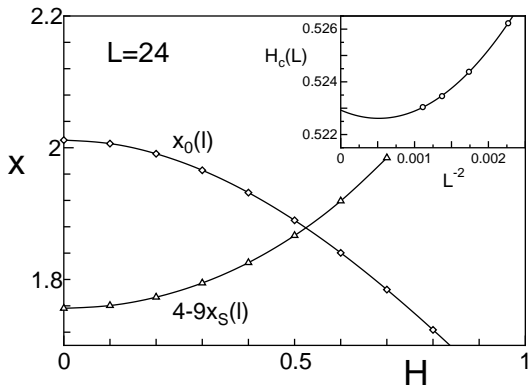}}
 \caption{
 The spectroscopy of levels observed in the $L=24$ sites system. 
 The crossing of $x_0(l)$ (diamonds) and $4-9x_S(l)$ (triangles) gives
 the finite-size estimate of the transition point $H_{\rm c}(L)$.
 Inset shows the extrapolation of finite-size estimates
 to the thermodynamic limit $L\to\infty$. 
 }
 \label{FIG2}
\end{figure}

\begin{table}[t]
 \caption{
 The $L$ dependences of dimensions $x_0(l)$ and $x_1(l)$ at $H_{\rm c}$.
 The average $x_{\rm av}(l)$ (see text) is extrapolated to $L\to\infty$
 using the least-squares fitting of the polynomial in $1/L^2$.
 } 
 \begin{tabular}{rcccccc}
  \hline\hline
    $L$~~ &18&21&24&27&30&$\infty$\\ 
  \tableline
  $    x_{     0}(l)$ &1.88528&1.88110&1.87913&1.87835&1.87823& ---    \\
  $    x_{     1}(l)$ &2.40464&2.37543&2.35551&2.34096&2.32981& ---    \\
  $    x_{\rm av}(l)$ &2.05840&2.04587&2.03792&2.03256&2.02875&2.01292 \\
  \hline\hline
 \end{tabular}
 \label{TAB_II}
\end{table}

 We perform the exact-diagonalization (ED) calculations of ${\bf
 T}_{Q_0}(L)$ for systems up to $L=30$.
 We show the level-crossing data in Fig.\ \ref{FIG2} and the
 extrapolation of the finite-size estimates $H_{\rm c}(L)$ to the
 thermodynamic limit $L\to\infty$ using the least-squares fitting of the
 polynomial in $1/L^2$ (the inset)
 \cite{Blot93,Card86}. 
 Then we find that while our result is consistent with previous
 estimations, it is much more accurate owing to the fast 
 convergence of finite-size estimates (see Table \ref{TAB_I}).
 Next we shall check a universal relation among excitation levels.
 For instance, the following relation is to be satisfied at the
 BKT point: $[2x_0(l)+x_1(l)]/3\simeq2$
 \cite{Nomu95}. 
 In Table\ \ref{TAB_II}, we give the scaling dimensions at $H_{\rm
 c}$, where the left-hand side of the relation is denoted as $x_{\rm
 av}(l)$.
 Although $x_0(l)$ and $x_1(l)$ considerably deviate from the value for
 the free-boson case 2 due to the logarithmic corrections, their main
 parts cancel each other, so the average takes a value close to 2.
 These data provide the check of the accuracy of our estimate and the
 evidences to ensure that numerically studied levels have the
 above-mentioned theoretical interpretations.

\begin{figure}[t]
\mbox{\includegraphics[width=2.8in]{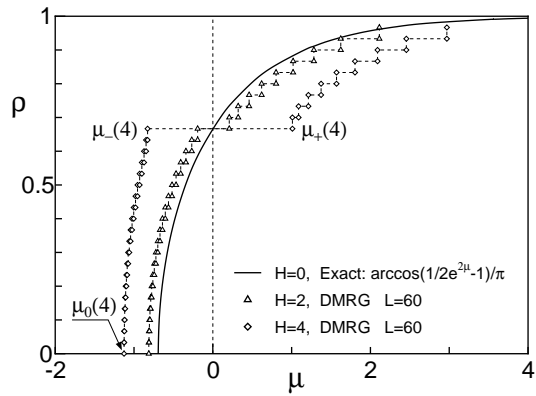}}
 \caption{
 The $\mu$-$\rho$ curves:
 Triangles (diamonds) with dotted line show the DMRG data at $H=2$
 ($H=4$), and the solid line exhibits the exact result at $H=0$. 
 The flat region with $\rho=\frac23$ corresponds to the string-density
 plateau $[\mu_-(H),\mu_+(H)]$.
 $\mu_0(H)$ is the threshold below which the string is absent. 
 }
 \label{FIG3}
\end{figure}

 In the rest part, we shall discuss the anisotropic case $\mu\ne0$.
 As we have seen in the above, the magnetic field $H$ favors the
 long-range ordered commensurate (C) phase through the potential
 $\cos 3\sqrt2\phi$.
 On the other hand, $\mu$ which newly introduces a local density term
 $\partial_1 \phi$ to the effective theory
 (\ref{eq_SG}) tends to stabilize an incommensurate (IC) liquid phase
 \cite{Pokr79,Schu80}.
 Therefore, as can be found in the literature, the PT-type C-IC
 transition may occur
 \cite{Noh_95,Blot82,Noh_94}.
 For $H=0$, we know the exact $\mu$-$\rho$ curve
 $\rho(\mu)=\arccos\left(1/2e^{2\mu}-1\right)/\pi$
 which is given in Fig.\ \ref{FIG3}
 \cite{Blot82}.
 On the other hand,
 for $H\ne0$, we employ numerical methods and
 estimate the curve from the finite-size system data as 
 $\mu\simeq\left[E_{{\rm g},Q+2}(L)-E_{{\rm g},Q}(L)\right]/2$
 \cite{Noh_95}.
 While, like the exact solid curve, $\rho$ is a smooth function of $\mu$
 showing the compressive liquid state for $H\le H_{\rm c}$, there is the
 string-density plateau $[\mu_-(H),\mu_+(H)]$ with $\rho=\frac23$ 
 for $H>H_{\rm c}$ (see the DMRG data denoted by marks).
 In this plot, the top and the bottom of each step correspond to
 $(Q+2)/L$ and $Q/L$ respectively, so we can estimate the C-IC phase
 boundary lines from the edges of the plateau. 

\begin{figure}[t]
\mbox{\includegraphics[width=2.8in]{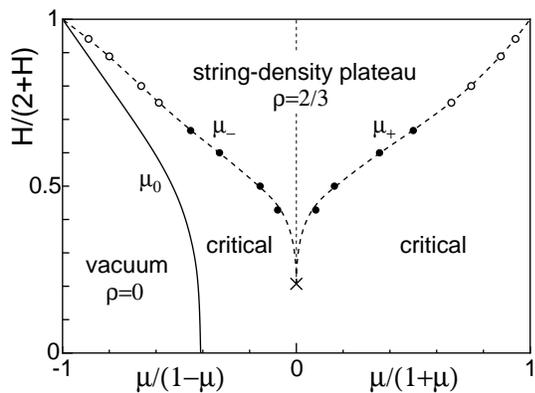}}
 \caption{
 The ground-state phase diagram.
 The vacuum of strings is stabilized on the left of the solid curve
 $\mu_0(H)$.
 The field-induced ordered phase corresponds to the string-density
 plateau state with $\rho=\frac23$.
 The cross on the $\mu=0$ line shows the BKT point obtained by the LS
 method.
 The filled (open) circles show estimates $\mu_\pm(H)$ by the DMRG (ED)
 method.
 Dotted curves give a guide to the eyes.
 }
 \label{FIG4}
\end{figure}

 In Fig.\ \ref{FIG4}, we provide our phase diagram.
 Here it is noted that the threshold $\mu_0(H)$ below which the
 doubly degenerate vacuum of strings with $\rho=0$ is realized (see
 Fig.\ \ref{FIG3}) is exactly given by $\mu_0(H)=-\ln[2\cosh(H/4)]$
 \cite{Blot82,Lin_79}.  
 So, we also draw it in the figure.
 The cross on the isotropic line $\mu=0$ shows the BKT point obtained by
 the LS method.
 For large $H$, we can use the ED data
 $\mu_\pm(H,L)=\pm\left[E_{{\rm g},Q_0\pm2}(L)-E_{{\rm g},Q_0}(L)\right]/2$ 
 and the extrapolation formula 
 $\mu_\pm(H,L)\simeq\mu_\pm(H)+{\rm const}/L^2$
 (see open circles)
 \cite{Saka91}.
 For $H\lesssim 4$, assuming the square-root behavior around the
 plateau, we estimate $\mu_\pm(H)$ from the $\rho$-$\mu$ curve obtained
 by DMRG.
 From this plot, we find that two PT-transition lines $\mu_\pm(H)$ seem
 to be terminated at the BKT point $(\mu,H)=(0,H_{\rm c})$ and that the
 plateau region becomes wider with the increase of the magnetic field. 
 For $H\simeq H_{\rm c}$, it is still difficult to determine the narrow
 plateau region corresponding to the exponentially small energy gap even
 by the use of the DMRG method.
 However, by combining the LS result and the DMRG data, we can obtain
 the reliable phase diagram of our model.

 Lastly, we shall discuss some related topics. 
 The magnetization process observed in the ground state of the
 $S=\frac12$ anisotropic AF chain with a strong frustration 
 exhibits the plateau at $\frac13$ of the saturation magnetization
 \cite{Okun02}.
 This plateau state exhibits spontaneous breaking of the translational
 symmetry down to the period $n=3$, so it is threefold degenerate (i.e.,
 $\uparrow\uparrow\downarrow\uparrow\uparrow\downarrow\!\cdots$), and
 thus satisfies the necessary condition for the magnetization plateau
 \cite{Oshi97}, i.e.,
 $n(S-m)={\rm integer}$ with the average magnetization per site
 $m=\frac16$.
 According to the bosonization treatment
 \cite{Lech04},
 this magnetization-plateau-formation transition is described by the
 sine-Gordon field theory, which is identical to the present case. 
 And more generally, since $Q$ is the conserved quantity in our system,
 the above necessary condition can be translated to the density plateau
 condition for the string systems as $n\rho={\rm integer}$, where $\rho$
 is the possible density at the plateau and $n$ is the periodicity of the
 plateau state.
 For its derivation, 
 ${\cal U}=\exp[-i\sum_{j_1=1}^{L}\left(2\pi j_1/L\right)N_{j_1,j_2}]$
 plays a role of the twist operator.
 While the present system only possesses the $\rho=\frac23$ plateau
 state with the spontaneous symmetry breaking of ${\cal T}$ down to the
 period $n=3$, Noh and Kim intensively investigated the interacting
 string/domain-wall systems based on TAFIM with spatially anisotropic
 further-neighbor couplings
 \cite{Noh_94}.
 They used the Bethe-ansatz method to diagonalize the transfer matrix,
 and found the $\rho=\frac12$ plateau phase.
 Furthermore the period $n=2$ of the plateau state has been suggested in
 that case.
 Therefore, we think that the above necessary condition for the plateau
 can provide a important viewpoint for the understandings of the interacting
 strings embedded in TAFIM with various extensions.

 One of the authors (H.O.) thanks 
 M. Nakamura and K. Nomura
 for stimulating discussions. 
 This work was supported by
 Grants-in-Aid from the Japan Society for the Promotion of Science.



\begin{references}
\bibitem{Comm1}
 The nonlinear potential $\cos 6\sqrt2\phi$ is highly irrelevant, so it
 was dropped in our argument.
\end{references}
\end{document}